# Quasi-nondegenerate pump-probe magnetooptical experiment in GaAs/AlGaAs heterostructure based on spectral filtration

M. Surýnek, L. Nádvorník*, E. Schmoranzerová, and P. Němec

*Faculty of Mathematics and Physics, Charles University, Ke Karlovu 3,121 16, Prague 2, Czech Republic*

**We report on a quasi-nondegenerate pump-probe technique that is based on spectral-filtration of femtosecond laser pulses by a pair of mutually-spectrally-disjunctive interference filters. This cost- and space-efficient approach can be used even in pump-probe microscopy where collinear propagation of pump and probe pulses is dictated by utilization of a microscopic objective. This technique solves the contradictory requirements on an efficient removal of pump photons from the probe beam, to achieve a good signal-to-noise ratio, simultaneously with a needed spectral proximity of the excitation and probing, which is essential for magnetooptical study of many material systems. Importantly, this spectral-filtration of 100 fs long laser pulses does not affect considerably the resulting time-resolution, which remains well below 500 fs. We demonstrate the practical applicability of this technique with close but distinct wavelengths of pump and probe pulses in spatially- and time-resolved spin-sensitive magnetooptical Kerr effect (MOKE) experiment in GaAs/AlGaAs heterostructure, where a high-mobility spin system is formed after optical injection of electrons at wavelengths close to MOKE resonance. In particular, we studied the time- and spatial-evolutions of charge-related (reflectivity) and spin-related (MOKE) signals. We revealed that they evolve in a similar but not exactly the same way which we attributed to interplay of several electron many-body effects in GaAs.**

## I. INTRODUCTION

The pump-probe technique is well established stroboscopic optical method allowing to measure ultrafast dynamical response of various materials [1]. Here, a short intensive laser pulse (pump) excites the investigated system and its relaxation towards an equilibrium is measured by a weaker time-delayed laser pulse (probe) in a reflection or transmission geometry. If this technique is combined with an optical microscope, it is possible to achieve relatively high spatial- and time-resolutions simultaneously [2]. The spatial-resolution of such pump-probe microscope is determined mainly by a numerical aperture of the objective and by a wavelength of the laser light [3]. The time-resolution is primarily given by a duration of the laser pulses. In a traditional form of degenerate pump-probe experiment, pump and probe pulses are obtained by an intensity splitting of the output from a pulsed laser, i.e., their wavelengths are exactly equal. This variety of pump-probe experiment is probably the most frequently used one not only due to its simplicity (and the resulting low cost) but also due to the fact that the signal-to-noise ratio in the measured data is determined predominantly by a performance of the laser itself, which is usually rather good. The most frequently employed detection scheme involves a chopper-induced intensity modulation of a pump beam and monitoring the pump-induced changes of a probe beam by a lock-in technique [4]. However, to achieve a good signal-to-noise ratio in this experiment, it is necessary to separate the signal due to probe photons, where the required information is encoded, from the signal due to (scattered) pump photons, which acts as a noise. In principle, this separation could be done electronically if both pump and probe beams are modulated at two distinct frequencies and a signal at the sum (or difference) frequency is measured by the lock-in amplifier. Nevertheless, much better signal-to-noise ratio is usually obtained if pump photons are prevented from reaching the photodetector at all. If pump and probe pulses have different angles of incidence [see Fig. 1(a)], this can be achieved quite straightforwardly by spatially filtering the probe beam after the sample. The situation is much more complicated if pump and probe beams are collinear, which is usually the case for experiments with a high spatial-resolution where microscopic objective lenses have to be used [see Fig. 1(b)]. In this case, one can separate the pump and probe photons by setting their polarization to mutually-orthogonal linear polarizations and placing a polarizer in front of the detector. However, this option is not compatible with magnetooptical experiments where the measured signal is encoded in the probe polarization rotation [3]. Similarly, it cannot be applied in a time-resolved research of molecular dynamics where a separation of pure population dynamics from molecular structural changes and rotational diffusion is achieved by setting the "magic angle" between the polarization planes of pump and probe photons [5,6]. Therefore, the most efficient solution is to use pump and probe pulses with a different wavelength and to remove scattered pump photons from the probe beam by placing a spectral filter, which is transparent only for probe photons, in front of the photodetector. Moreover, if excitation and/or detection channels are strongly wavelength dependent, which is the case for magnetooptical effects [3], the independent spectral tunability of pump and probe pulses is very favorable also for the experimental setup performance optimization.

There are several well established methods how to obtain different wavelengths of pump and probe pulses for nondegenerate pump-probe experiment. In principle, this can be achieved by an electronic synchronization of the pulse trains emitted from two independent laser systems [7,8]. However, due to a presence of the timing jitter in the lasers, the achievable time-resolution in this case is limited to several picoseconds only, unless a rather complicated detection scheme is used [9]. In order to reach a femtosecond resolution, it is necessary to start with a single femtosecond laser which output is splitted into two (or more) parts. Using this approach, the simplest way how to change the wavelength of one of the beams is to use a second harmonic generation (SHG) in a nonlinear crystal. This method serves its purpose very well but only when the very different wavelengths of pump and probe pulses (e.g., in a case of Ti:sapphire laser, around 400 and 800 nm, respectively) and/or the inability to tune independently the wavelengths of the beams are not an issue. Considerably more versatile solution, which enables to achieve a spectral tunability in a very broad range, is to use one of the beams to pump an optical parametric oscillator (OPO) or optical parametric amplifier (OPA) [2,10]. The disadvantage of this solution is a high cost and a large footprint of OPO/OPA. To avoid this, very efficient light color conversion can be obtained also in a photonic crystal fiber [11]. However, as several distinct nonlinear effects take place simultaneously in photonic fibers, the achievable time-resolution in this case is limited to several picoseconds [11]. Moreover, because nonlinear optical effects are used in a photonic crystal fiber, OPO and OPA, in all these cases the resulting signal-to-noise ratio is usually considerably worse than the one obtained when the fundamental frequency output from the laser is used in the degenerate pump-probe experiment.

Another convenient option to obtain synchronized pump and probe pulses at different wavelengths is to select two distinct spectral regions within a spectrum of the femtosecond laser pulse. In reality, this seemingly straightforward solution is considerably complicated by the fact that spectral- and temporal-profiles of light pulses are mutually interconnected through the Fourier transform [12]. Selecting a rectangular spectral-profile from a laser pulse (by a monochromator, for example) leads to a *sinc$^2$* time-profile and, consequently, to a very poor time-resolution. To avoid this, more complicated approaches based on a grating-based pulse shaper [13,14] or a prism-pair [2] were used. But even here the experimentally demonstrated time-resolutions were in the picosecond time scale only [13,14]. In this paper we show that sub-picosecond time resolution can be achieved in a very simple quasi-nondegenerate pump-probe experiment where an output of a femtosecond oscillator is filtered by a pair of well-chosen mutually-spectrally-disjunctive interference filters. We also demonstrate that this cost- and space-efficient solution is very useful not only in the case of pump-probe microscopy, as we have already reported recently [15], but also in a standard time resolved spin-sensitive experiments in semiconductors.

The paper is organized as follows: First, we describe briefly experimental setups where this experimental technique was tested. Then we study spectral- and temporal behavior of the filtered femtosecond laser pulses. Finally, we demonstrate the utilization of this technique for a measurement of electron spin dynamics at the heterointerface GaAs/AlGaAs.

## II. EXPERIMENTAL SETUP

As a light source we used a computer-controlled femtosecond Ti:sapphire oscillator (Mai Tai, Spectra Physics) with a repetition rate of 80 MHz that generates ≈ 100 fs laser pulses with a spectral width of about 8 nm (full width at half maximum, FWHM). Each laser pulse is splitted to two parts with a typical intensity ratio between pump and probe pulses equal to 10:1. The intensity of the pump beam is modulated by a chopper at frequency of ≈ 2 kHz. A time delay between pump and probe pulses, $\Delta t$, is set by a computer-controlled delay line. The sample is mounted in an optical cryostat, where temperature was set to 15 K, and placed between poles of an electromagnet providing in-plane magnetic field up to 500 mT. Light reflected from the sample is led to a detection part of the setup. Here the magnetooptical (MO) signal corresponding to the probe polarization rotation and the differential reflectivity are measured as difference and sum of signals from detectors in the optical bridge, respectively [16]. Two possible implementations of the pump-probe setup are schematically depicted in Fig. 1. In Fig. 1(a) we show the version with a non-collinear incidence of pump and probe beams - the angles of incidence to the sample surface are < 1° and ≈ 7° for the pump and probe beams, respectively. Both beams are focused by a single converging 10D lens on the sample to overlapped spots with a size of ≈ 25 μm (FWHM). The major advantage of this configuration is that it enables, even without any spectral-filtering described in this paper, to filter out spatially the scattered pump photons from the probe beam, which leads to a very good signal-to-noise ratio in the measured data [17-19]. If higher spatial-resolution is needed, a microscopic objective lens has to be used instead of the focusing lens. However, the correct functionality of the objective lens requires pump and probe beams to be collinear [see Fig. 1(b)] and, consequently, it is not possible to separate pump and probe beams by a spatial-filtration. This in turn leads to a very bad signal-to-noise ratio unless other means of separating pump and probe photons are implemented. As we show below, a very compact and cost-efficient solution of this problem, which *does not* affect significantly the achievable time-resolution, is to use pump and probe beams prepared from the output of a femtosecond oscillator by a spectral-filtering using a pair of mutually-spectrally-disjunctive interference filters. The resulting quasi-nondegenerate experiment can be used very efficiently in experiments with a high spatial-resolution; see Ref. 15 where a near-infrared objective with magnification 20x and numerical aperture 0.4 was used to create a spot size < 2 μm. In addition, as we demonstrate below, this rather small variation of the experimental setup can be used to enhance considerably the measured time resolved MO signal in a case of materials with spectrally narrow MO spectrum, or even to measure this MO spectrum in a pump-probe experiment.

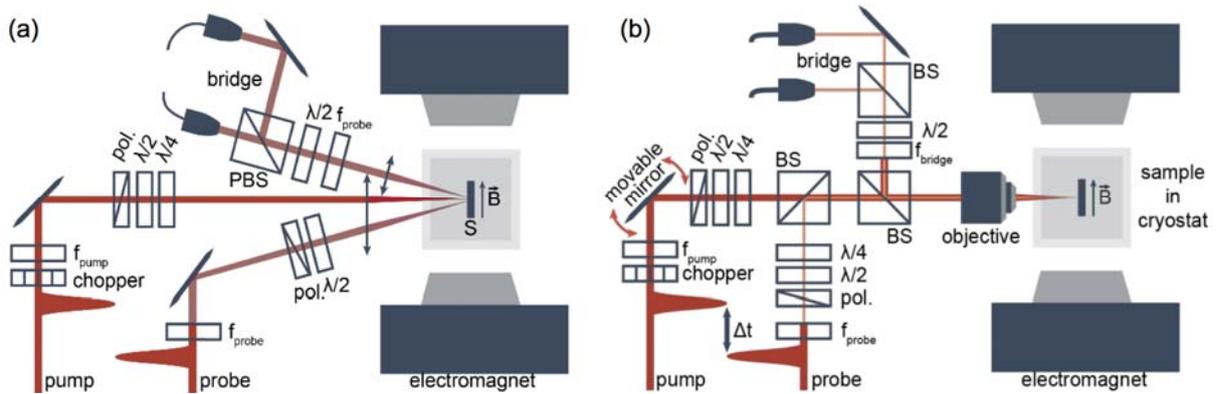

**Figure 1: A schematic depiction of the experimental pump-probe setups where spectral-filtering was tested.** (a) Non-collinear incidence of pump and probe beams used in experiments where light was focused by a converging lens to a spot size of ≈ 25 μm. (b) Collinear incidence of pump and probe beams used in experiments with a microscopic objective, which was focusing light to a spot size < 2 μm. Polarization of beams was controlled by a combination of a polarizer (pol.), half-wave plate ($\lambda/2$) and quarter-wave plate ($\lambda/4$). Beams were combined and/or split by a non-polarizing (BS) and polarizing (PBS) beam splitters. The interference filters for the spectral-filtration of pump and probe pulses are marked as $f_{pump}$ and $f_{probe}$, respectively.

## III. SPECTRAL FILTRATION OF FEMTOSECOND LASER PULSES

In our experiment, we used a rather typical commercial femtosecond Ti:sapphire oscillator which produces ≈ 100 fs laser pulses with a central wavelength tunable in the infrared spectral region. The major problem connected with a spectral filtration of these femtosecond pulses is their relatively narrow spectrum with a width of about 8 nm (FWHM). To prepare a pair of mutually-spectrally-disjunctive laser pulses, it is necessary to use pairs of spectral filters with a sufficiently separated transmission edges (see Fig. 2). Moreover, as already discussed in the Introduction, the filter transmission edges could not be too steep, otherwise the mutual interconnection between the spectral- and temporal-profiles of light pulses would lead to a considerable prolongation of the filtered laser pulse in a time-domain. Simultaneously with this, the transmission edges have to be rather steep and located relatively close to each other (i.e., within the spectrum of the oscillator output) otherwise the intensity of transmitted laser pulses would not be sufficient for the pump-probe experiment. Several combinations of commercially available interference filters (bandpass, notch, long pass, and short pass) sold by Thorlabs, which are matching these mutually antagonistic criteria, are listed in Table 1 and their spectral dependences of transmittances are shown in Fig. 2 as dashed lines. The spectra of transmitted laser pulses are dependent also on the spectral profile of pulses that are incident on the filters. The filled curves in Fig. 2 depict the spectral profiles of the unfiltered oscillator outputs at the sample position for several selected central wavelengths. Apparently, at certain wavelengths, the propagation in air (for about 5 m) and/or properties of optical components in the beam path modify partially the pulse spectrum from the Gaussian shape of transform-limited laser pulses, which are leaving the oscillator cavity. The solid lines in Fig. 2 represent the measured spectra after insertion of the corresponding filters to positions depicted in Fig. 1(a). Note that for a given filter pair, the relative intensity of the transmitted pulses in the pump and probe arms of the setup, respectively, can be fine-tuned by slightly changing the central wavelength of the oscillator output.

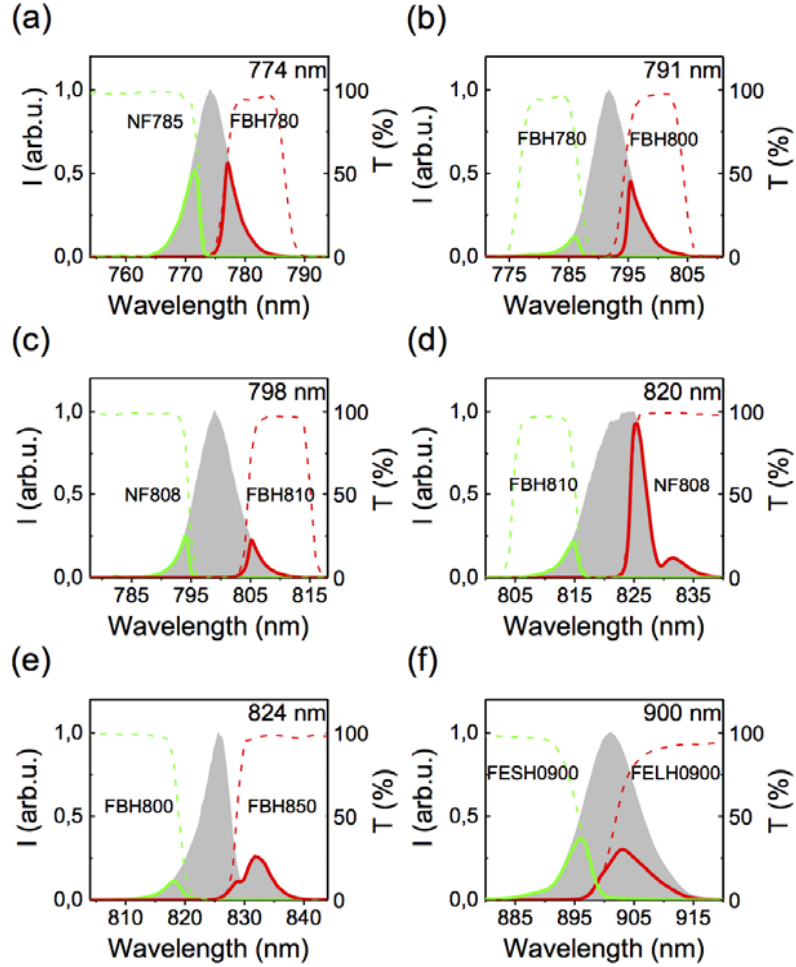

**Figure 2: Examples of spectrally-disjunctive laser pulses (solid lines) obtained by filtering femtosecond laser pulses** from the oscillator (filled curves) with central wavelengths of (a) 774 nm, (b) 791 nm, (c) 798 nm, (d) 820 nm, (e) 824 nm, and (f) 900 nm. The dashed lines show transmission spectra of the corresponding interference filters.

We stress that the ability to create a pair of spectrally-disjunctive laser pulses by spectrally filtering an incident femtosecond laser pulse is not a guarantee of their applicability in the pump-probe experiment. The interference filters are rather complicated layered structures that can, in principle, distort severally the time profile of the laser pulses. To address this issue, we measured the intensity auto- and cross-correlations [12] of the pulses as described below. We used the setup shown in Fig. 1(a) where we replaced the sample by a beta barium borate (BBO) nonlinear crystal. We detected (in transmission geometry) light produced by SHG when pump and probe pulses, which were set to have the same intensity, were present in the BBO crystal simultaneously. The filled curves in Fig. 3 correspond to the correlation traces of the unfiltered oscillator outputs for several selected central wavelengths. As expected, these curves are symmetrical with respect to a zero-time delay, which is a signature that properties of pump and probe pulses in our experimental setup are rather similar – i.e., these curves can be regarded as auto-correlation traces for the oscillator outputs. The 3-peak structure observed for the central wavelength of 900 nm indicates that at this wavelength there is a weaker satellite with a time-spacing of about 0.5 ps relative to a position of the major laser pulse. The widths (FWHM) of the measured auto-correlation traces, which correspond to an actual time-resolution in our pump-probe experiment, are listed in Table 1. The precise shape of the laser pulses envelope is not known in our case. If we assume *sech²*-shaped pulses for simplicity, the pulse duration is ≈ 0.65 times the width of the measured autocorrelation signal, which corresponds to 80 – 120 fs for the wavelengths studied. Next, we inserted the selected filters in one of the beams and we measured the corresponding correlation trace again. For certain filters, the obtained correlation trace remained nearly symmetrical with respect to a zero-time delay, which shows that the time-profile of the filtered laser pulse was not modified significantly by the filter [see Fig. 3(e)]. However, for a majority of the filters the measured correlation trace is clearly asymmetrical in time, which can be interpreted as a broadening of the trailing edge of the laser pulse in the interference filter. Nevertheless, for all used filters the achievable time resolution in the corresponding pump probe experiment is well below 500 fs (see Table1).

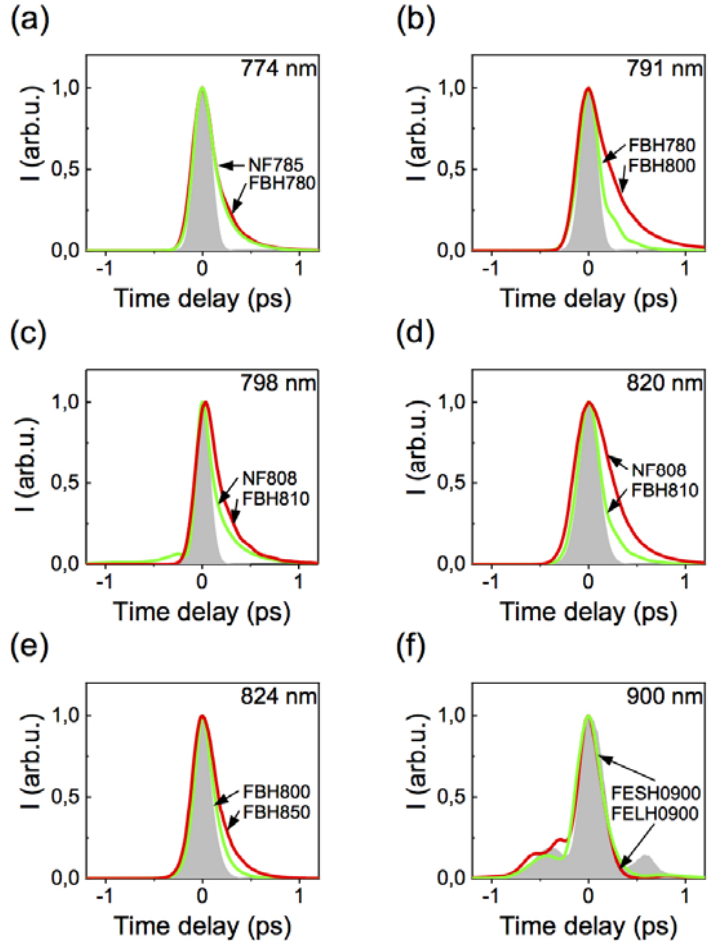

**Figure 3:** Correlation traces measured by non-collinear SHG in BBO crystal for unfiltered (filled curves) and filtered (solid lines) laser pulses.

Last but not least, we would like to mention one additional advantage of this filter-based method. Unlike in the case of OPO and OPA, the spectral position of the obtained laser pulses does not fluctuate in time because the position of its spectrally-sharp edge is set by a position of the filter transmission edge (see Fig. 2). This is very important in all experiments when "seemingly small" thermal spectral fluctuations of laser pulses (e.g., for about 1 nm) could lead to a large change of detected signals. As an example, we can mention magnetooptical experiments in the field of semiconductor spintronic [20-23]. Here, imaging of the lateral spin transport is based on magnetooptical Kerr effect (MOKE) the spectrum of which is usually rather narrow and oscillatory [22], see also Fig. 4(e).

| $\lambda_{central}$ (nm) | Filter | $\lambda_{max}$ (nm) | $I/I_0$ (%) | $\Delta t_{FWHM}$ (fs) |
|---|---|---|---|---|
| 774 nm | – | – | – | 126 |
| | FBH780-10 | 777 | 24.6 | 204 |
| | NF785 | 772 | 25.3 | 220 |
| 791 nm | – | – | – | 124 |
| | FBH780-10 | 786 | 5.9 | 206 |
| | FBH800-10 | 795 | 19.8 | 240 |
| 798 nm | – | – | – | 114 |
| | NF808 | 794 | 8.8 | 230 |
| | FBH810-10 | 805 | 9.7 | 215 |
| 820 nm | – | – | – | 145 |
| | FBH810-10 | 815 | 7.3 | 193 |
| | NF808 | 826 | 31.2 | 390 |
| 824 nm | – | – | – | 121 |
| | FBH800-40 | 818 | 6.4 | 180 |
| | FBH850-40 | 832 | 20.9 | 172 |
| 900 nm | – | – | – | 189 |
| | FELH0900 | 903 | 46.5 | 272 |
| | FESH0900 | 896 | 19.5 | 241 |

**Table 1: Summary of properties of original and filtered femtosecond laser pulses.** For each oscillator output, which is labelled by a corresponding central wavelength ($\lambda_{central}$), the properties of laser pulses after passing through a given interference filter are shown: spectral position of maximal intensity ($\lambda_{max}$), transmitted fraction of the laser pulse ($I/I_0$), temporal length of correlation trace ($\Delta t_{FWHM}$) measured by SHG. The filters are labelled by the notation of Thorlabs: Bandpass filters are denoted as "FB", notch filters as "NF", long pass filters as "FEL", and short pass filters as "FES"; the letter "H" depicts that the filter is a premium version of a filter with improved properties (higher transmission, steeper cut-on and cut-off slopes). The number behind the letters describes a spectral position of the filter and the second number for bandpass filters stands for their spectral width (in nanometers).

## IV. QUASI-NONDEGENERATE PUMP-PROBE EXPERIMENT IN GAAS/ALGAAS HETEROSTRUCTURE

In this chapter we demonstrate the practical applicability of our quasi-nondegenerate pump-probe experiment for an investigation of the dynamics of spin-polarized electrons in semiconductors. The studied sample was a simple undoped heterostructure that is schematically depicted in Fig 4(a). A 100-nm-thick undoped $Al_{0.4}Ga_{0.6}As$ barrier was deposited by molecular beam epitaxy on top of an insulating GaAs buffer and a GaAs substrate. The barrier was then covered by another undoped 800-nm-thick GaAs layer where the confined long-lived and highly mobile electron subsystem was formed as a consequence of the spatial separation of optically generated electron-hole pairs in a built-in electric field due to surface states (see Refs. 15 and 24 for details).

We used the optical orientation effect to generate spin-polarized electrons by absorption of circularly-polarized pump pulses [25]. This effect produces ≈ 50% spin polarization of photoinjected electrons in a semiconductor but only if the pump photon energy is between the band gap energy ($E_g$ = 1.519 eV for GaAs at low temperatures) and the energy corresponding to the transition between a conduction band and a spin-orbit splitted valence band [25]. Outside this energetical window, which has a width of ≈ 340 meV for GaAs (see Fig. 2 in Ref. 25), circularly polarized pump pulses do not produce any sizable spin polarization. Similarly, the MO coefficient is non-zero only around $E_g$ in non-magnetic semiconductors [Ref. 22, see also Fig. 4(e)], which restricts considerably the possible wavelengths of probe pulses. We stress that these two above mentioned factors limit the usable wavelengths of pump and probe beams not only for our particular samples. The requirement of a spectral proximity of pump pulses, probe pulses and band gap energy is a common feature for all spin-sensitive pump-probe experiments in any non-magnetic semiconductor.

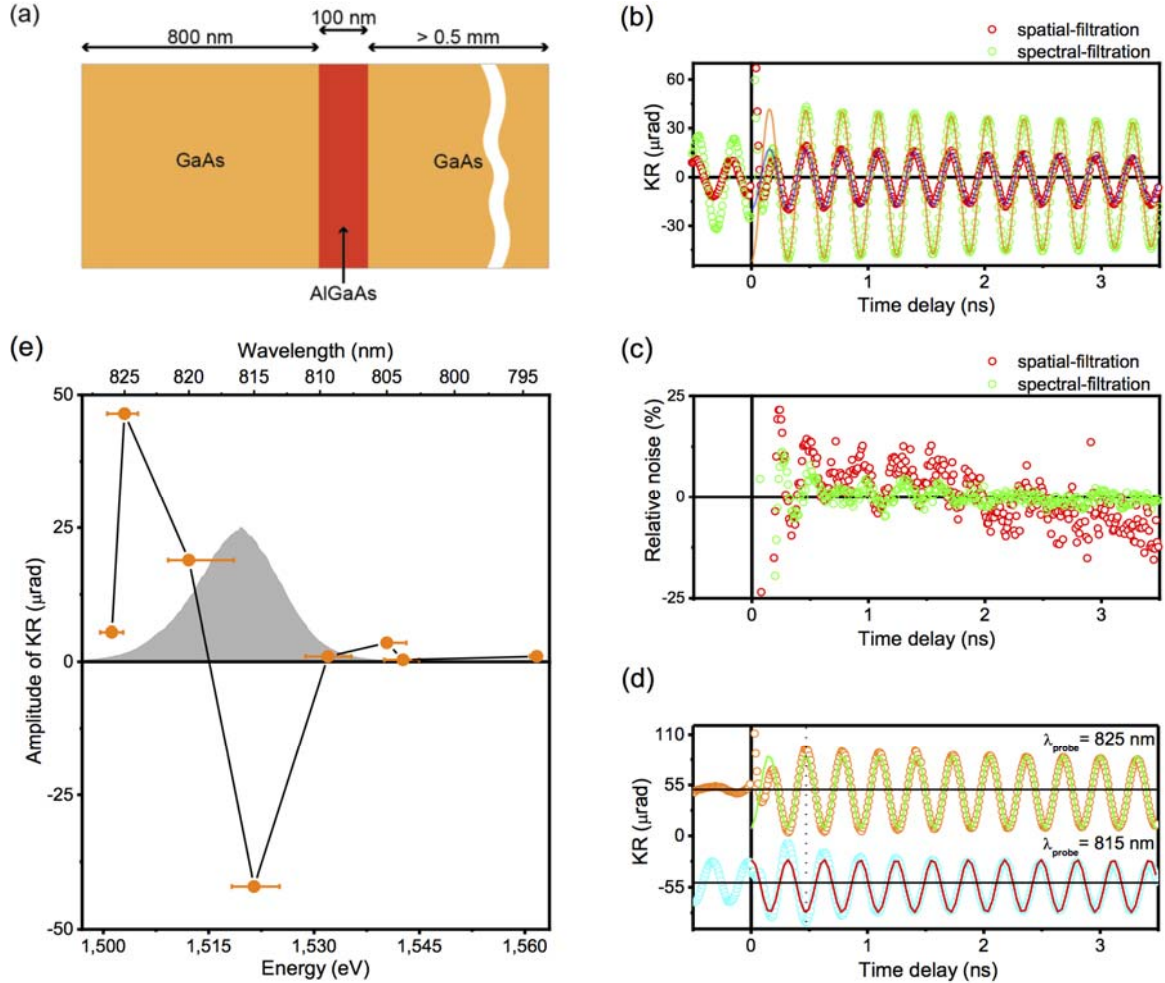

**Figure 4: Utilization of quasi-nondegenerate pump-probe experiment for an investigation of the dynamics of spin-polarized electrons in GaAs/AlGaAs heterostructure**. (a) Sketch of the layer structure of the studied sample (the sample surface, on which the light is incident, is on the left side). The studied long-lived and highly mobile electron spin-system is self-consistently formed near the upper GaAs/AlGaAs interface. (b) Time resolved polarization rotations due to MOKE that were measured using a spatial-filtration by laser pulses with central wavelength of 816 nm (red points) and the same pulses that were spectrally-filtered (by FBH810-10 and FB820-10 in pump and probe beams, respectively) (green points) for external magnetic field of 500 mT applied in the sample plane at 10 K. In both cases, the fluence of pump pulses was 8 µJ.cm$^{-2}$. The solid lines are fits by a damped harmonic function. (c) Comparison of the relative noise level in data shown in part (b); the noise, computed as a difference between the experimentally measured data and the theoretical fit, was divided by the measured data amplitude. (d) Demonstration of MO signal phase change measured in a quasi-nondegenerate pump-probe experiment due to probe pulses wavelength change. The vertical line depicts the time delay of 470 ps where the polarization rotations displayed in part (e) were recorded. (e) Spectral shape of MOKE in the studied electron sub-system that was derived from the measured time resolved MO signals in the quasi-nondegenerate experiments. The position and horizontal bar width for each data point correspond to the maximum and spectral width of the spectrally-filtered probe pulses, respectively. The filled curve depicts the unfiltered spectrum of laser pulses that was used to measure the data shown in part (b).

For initial test experiment we used the experimental setup with a non-collinear incidence of pump and probe beams [see Fig. 1(a)] where the separation of pump photons from probe beam can be achieved both by a spatial- and spectral-filtration of the probe beam reflected from the sample. (Note that if we used the setup with a collinear incidence of pump and probe beams, Fig. 1(b), the pump-probe signals without the spectral-filtration *could not* be measured at all because it would be completely masked by scattered pump photons.) In Fig. 4(b) we show the experimental data measured by these two filtration methods, which demonstrate the very slow damping of the electron spin precession in the studied sample [15,24]. Interestingly, the data measured by the spectral-filtration have a larger amplitude of the measured MO signal. This is a direct consequence of a narrower bandwidth of the spectrally-filtered pulses that are polarization-rotated in the sample by magnetooptical Kerr effect (MOKE), whose spectrum is usually rather narrow and oscillatory in semiconductors (see the inset in Fig. 1 in Ref. 22). Consequently, if spectrally-broad laser pulses are used, which is the case for unfiltered 100 fs long laser pulses, the measured dynamical MO signals can be considerably reduced if the high- and low-energy parts of the laser pulse experience positive and negative polarization rotations, respectively [see Fig. 4(e)]. Using the well-defined precession nature of the studied spin-sensitive signal in an external magnetic field, which is represented by the theoretical fit in Fig. 4(b), we can evaluate the relative noise level in the measured data, which contain both the signal and noise, as depicted in Fig. 4(c). Clearly, there is a considerably lower noise level in the measured data when the spatial-filtration is applied, which is a direct consequence of a very good separation of pump and probe photons by the used combination of interference filters.

If the wavelength of probe pulses is changed, not only the precession signal amplitude but also its phase can be changed – see Fig. 4(d). This strong sensitivity of the amplitude and phase of the measured dynamical MO signal on the relative spectral positions of probe laser pulses and MOKE spectrum can be used for a crude reconstruction of the MOKE spectrum in the studied sample from the measured time resolved data, as demonstrated in Fig. 4(e). This procedure is extremely useful in the case of the studied material system where the observed long-range and high-speed electronic spin-transport is present only in a very thin layer of GaAs close to the GaAs/AlGaAs interface, which is formed by a built-in electric field that is separating the photo-generated electrons and holes [15,24]. The standard (time-unresolved) experimental techniques for measuring the MOKE spectrum [26] would provide information only about the spatially-averaged magnetooptical properties of the sample, which would be the one corresponding to undoped GaAs in our case. On the contrary, the MOKE spectrum shown in Fig. 4(e) corresponds solely to photo-injected electrons with a very long spin lifetime, which reside in the triangular quantum-well-like states close to the GaAs/AlGaAs interface (see Fig. 2(b) in Ref. 15). The absolute position of the MOKE spectrum is strongly dependent on the stress-induced band edge shifts in GaAs [22] and, therefore, its comparison is rather difficult between different experiments. The most apparent distinction between the MOKE spectrum in our material system, when electrons are injected to GaAs optically, and the n-doped GaAs epilayer [22] is the spectral width – the MO coefficient is non-zero only within ≈ 15 meV for the latter material system (see Fig. 2(d) in Ref. 22) while in our case the spectrum is considerably broader [see Fig. 4(e)].

For a final test experiment we used the experimental setup with a collinear incidence of pump and probe beams. The simultaneous high spatial- and time-resolutions of our experimental technique allows us to study exotic spin-transport phenomena in the high-mobility regime of optically excited GaAs/AlGaAs interface. An example of a contra-intuitive behavior of spin diffusion is shown in Fig. 5 where the relative positioning of pump and probe beams on the sample surface was controlled by a movable mirror [see Fig. 1(b)]. As pointed out in Ref. 27, the interpretation of MO experiments in terms of spin dynamics is truly meaningful only past the temporal overlap between pump and probe pulses and/or after the characteristic electronic dephasing time in the studied material. Therefore, the utilization of "as short as possible" laser pulses is essential for investigation of the initial stages dynamics of photo-injected carriers and spins. The measured spatially-resolved traces of Kerr rotation and transient reflectivity for several selected time-delays $\Delta t$ are shown in the left and right panels of Fig. 5(a), respectively. We note that is necessary to measure both these quantities for disentangling the dynamics of spin and charge [28]. We observed that a few picoseconds after the photo-excitation of the sample by a pump pulse with a Gaussian spatial profile both the MOKE and transient reflectivity signals show the expected Gaussian-like distribution of spins and carriers [see the traces for $\Delta t$ = 5 ps in Fig. 5(a)]. However, after another 10 ps, the profiles of both signals change dramatically and considerable modulation and even a sign change occur in the measured traces. We also note that the dynamics of spin (MOKE) and charge (transient reflectivity) is apparently rather different – see, e.g., the traces for $\Delta t$ = 95 ps in Fig. 5(a).

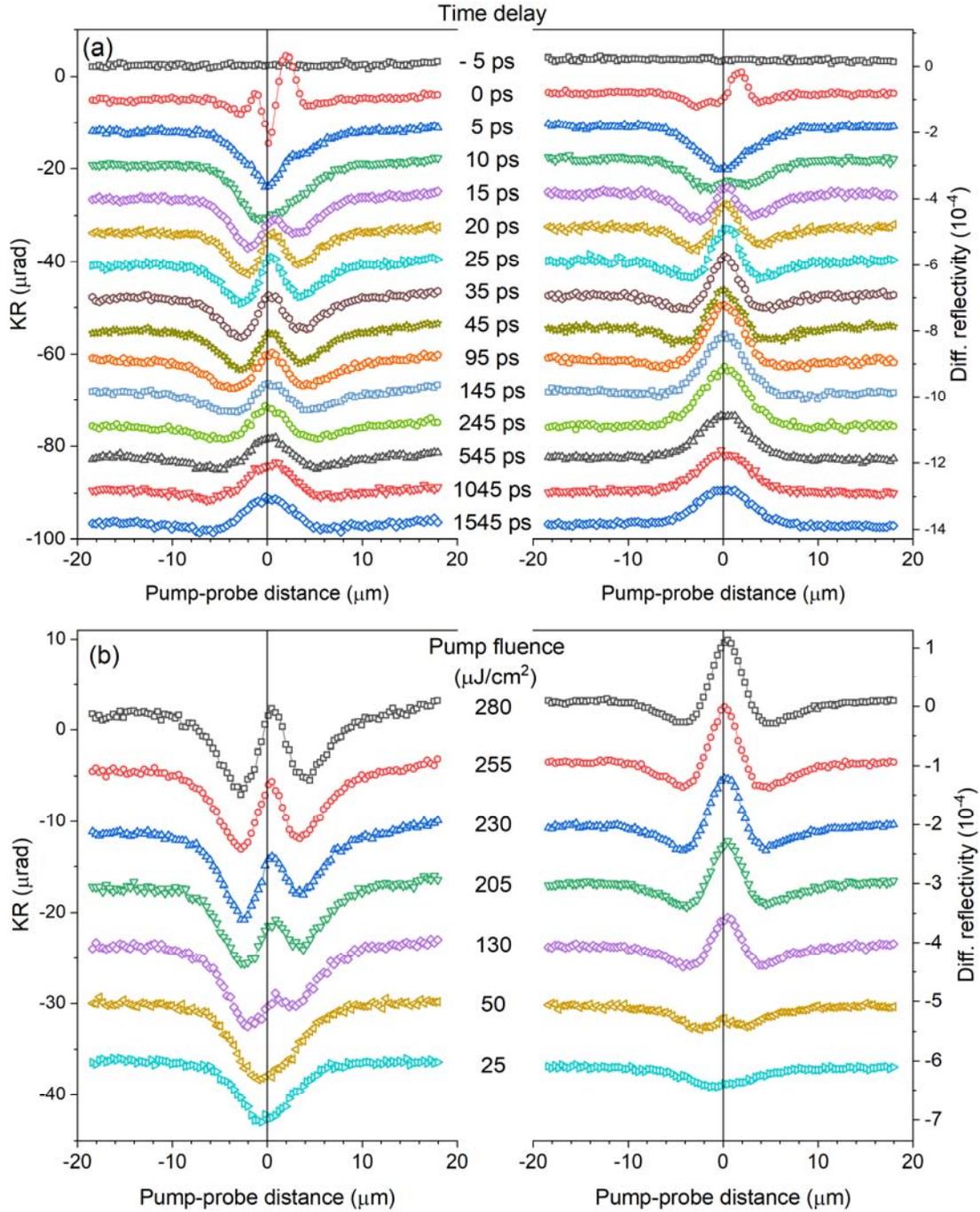

**Figure 5: Utilization of quasi-nondegenerate pump-probe experiment for an investigation of the time- and spatially-resolved spin and charge diffusion in GaAs/AlGaAs heterostructure** using the experimental setup with a collinear incidence of pump and probe beams. (a) Kerr rotation (left panel) and differential reflectivity signals (right panel) as functions of pump-probe spot distance for various pump-probe time delays (central labels) for pump power 305 µJ.cm-2. (b) Same signals measured at time delay 45 ps for various pump powers. Filters NF808 and FBH810-10 [see Fig. 2(d)] were used in pump and probe beams, respectively. Measured curves are vertically shifted for clarity.

The observed spatial evolution of the (spin-sensitive) MOKE signal for time delays < 100 ps shares rather prominent similarities with the spatially-resolved time-integrated photoluminescence (PL) measured under comparable conditions in p-doped GaAs [29-31]. Similar to our data, there was a dip in the PL polarization profile observed at a center of the photo-excitation spot if degeneracy of the electron gas was achieved (i.e., for a high concentration of electrons at low temperature). The authors attribute it to the Pauli blockade and the spin-degeneracy pressure if the photoexcited carrier density surpasses the steady-state density. In this high-degeneracy regime, the scattering probability is reduced due to the higher occupation of end-states of the scattering process. When the photocarriers are spin-polarized, it leads to different mobility and diffusion constants and, thus, different diffusion profile for minority and majority spin populations which forms a characteristic dip at the center of excitation. The interpretation of the observed oscillatory behavior in the measured traces as a consequence of electron many-body effects in GaAs [29-32] is corroborated by the fact that these effects are apparent only at high excitation intensities [see Fig 5(b)]. As our interfacial system features much longer spin life-time (10's of ns comparing to ≈ 200 ps in Ref. 29), we can observe the evolution of the system up to nanosecond time scale where a complete inversion of the MOKE signal is found [see Fig.5(a)]. However, this interpretation does not explain the evolution of the transient reflectivity signal, which is shown in Fig.5(b), where we observed similar trends but time-shifted (accelerated) with respect to the MOKE signal. The reflectivity signal is connected with a carrier-induced change of complex index of refraction, which is sensitive to the density of photo-injected charge-carriers and their energetical position within the semiconductor band structure. Within the Pauli blockade picture, the difference in the diffusion profiles for both spin populations should lead to simple Gaussian profile in the charge-carrier distribution. In previous experiments, non-Gaussian ring-like features were reported in spatially-resolved PL experiments in double [33-35] and single GaAs-based quantum wells [36] and were explained in terms of indirect excitons and a Coulomb interplay of cold and hot photoexcited electrons and holes, respectively. Another explanation for a similar unusual charge diffusion was suggested in novel semiconducting materials, such as transition metal dichalcogenides, where it was attributed to the Auger recombination [37,38].

As our experiment shows, the time- and spatial-evolutions of charge-related (reflectivity) and spin-related (MOKE) signals evolve in a similar but not exactly the same way. Consequently, none of the proposed purely spin- or charge-related effects can explain our experimental observations entirely and their combination and/or involvement of other mechanisms, such as Coulomb spin drag [39,40], should be considered. Although the precise identification of the exact underlying effect(s) is beyond the scope of this article, the reported experiment clearly demonstrates the very good applicability of our experimental method for a research of the mutual inter-coupling between the spin and charge transport dynamics in a regime with a simultaneously long spin life time and large diffusion constant.

## V. CONCLUSIONS

We demonstrated that it is very simple to transform a standard degenerate pump-probe MO experiment with ≈ 100 fs laser pulses into its quasi-nondegenerate form using spectral-filtration of the laser pulses by appropriately-selected interference filters. The major advantage of this approach is that it is cost- and space-efficient and, simultaneously, it does not affect considerably the resulting time-resolution, which remains well below 500 fs. The utilization of this approach leads to the improved signal to noise ratio in the pump-probe experiment and it enables to use an arbitrary polarization of pump and probe pulses, because pump and probe photons do not have to be separated by polarization means. The full potential of this technique lies in pump-probe microscopy where collinear propagation of pump and probe pulses is dictated by the utilization of a microscopic objective and where, consequently, the possibility to remove scattered pump photons from the probe beam by a spectral-filtration is essential for achieving a good signal-to-noise ratio. To demonstrate the capability of our experimental technique for pump-probe microscopy we studied the dynamics of spin and charge in highly mobile self-confined electron subsystem formed at the undoped GaAs/AlGaAs heterointerface. We observed a rather contra-intuitive behavior of the spin and carrier diffusion which we attributed to interplay of several electron many-body effects in GaAs.

**ACKNOWLEDGMENTS**

This work was supported by the Grant Agency of the Czech Republic under EXPRO Grant No. 19-28375X, by EU FET Open RIA under Grant No. 766566, and from the Charles University Grants No. 1582417 and No. SVV-2019-260445.